\documentclass[aps,epsfig,preprint]{revtex4-1}
\usepackage[section]{placeins}        
\usepackage{float}
 \usepackage[T1]{fontenc} 
 \usepackage{ae,aecompl}
 \usepackage{graphics}
 \usepackage{psfrag}
 \usepackage{multirow}
\usepackage{caption}
\usepackage{epsfig}
\usepackage{pdfpages}
\usepackage{subcaption}\usepackage{ragged2e}
\usepackage{amsmath,amsthm, amssymb, latexsym}

\newcommand{\be}{\begin{equation}}
\newcommand{\ee}{\end{equation}}

\newcommand{\bea}{\begin{eqnarray}}
\newcommand{\eea}{\end{eqnarray}}

\begin{document}
\title{ Particle-in-Cell simulation of two-dimensional electron velocity shear driven
instability in relativistic domain  }
 \author{Chandrasekhar Shukla}
 \email{chandrasekhar.shukla@gmail.com}
 \author{ Amita Das}
 \email{amita@ipr.res.in}
 \affiliation{Institute for Plasma Research, Bhat , Gandhinagar - 382428, India }
  \author{Kartik patel} 
 \affiliation{Bhabha Atomic Research Centre, Trombay, Mumbai - 400 085, India }
 
\date{\today}
\begin{abstract} 
 We carry out Particle-in-Cell (PIC) simulations to study the instabilities associated with a 2-D sheared electron flow configuration  against a neutralizing background of ions. Both weak and strong 
 relativistic flow velocities are considered.  
In the weakly relativistic case, we observe the development of 
 electromagnetic Kelvin Helmholtz instability with 
similar characteristics as  that predicted by the  
 electron Magnetohydrodynamic (EMHD) model. On other hand, in strong relativistic case the compressibility effects of electron fluid dominate and introduce upper hybrid electrostatic oscillations transverse to the flow which 
 are very distinct from EMHD fluid behaviour. In the nonlinear regime, both weak and strong relativistic cases lead to turbulence with broad power law spectrum. 
 
  \end{abstract}
 \maketitle 
\section{Introduction}
 The fundamental physical processes which governs the evolution of electron flows with velocity gradient are of great interest in wide range of research areas in astrophysical and laboratory contexts. 
In astrophysical scenario, the relativistic jets which are observed across wide range of astrophysical scales from micro-quasars to Gamma Ray Bursts (GRBs), supernovas etc., \cite{J.H.N,GRB,GRB1,SUPERNOVA} would have sheared flow of electrons. 
In laser plasma experiments also, there are many situations where the sheared electron flow configuration is inevitable. For instance 
experiments on fast-ignition scheme of laser-driven inertial confinement fusion involve electron beam propagation inside a plasma  which would invariably 
result in a sheared configuration of electron flow. When a high intensity laser irradiates a solid surface and/or a compressed plasma it generates 
 electron beam at the critical density surface of the plasma by the wave breaking mechanism \cite{malka,Modena,joshi}. This beam typically propagates inside the high density region of the plasma exciting reverse shielding background 
 electron currents. The forward and reverse currents spatially separate 
 by Weibel instability leading to a sheared electron flow configuration. However, since   the transverse extent of the  beam is finite compared to the plasma width,  being 
  commensurate with the laser focal spot, the sheared configuration of  electron flow automatically  exists between the beam and the background 
 stationary  electrons  at the edge of the propagating beam \cite{chandra} even before 
 Weibel destabilization process. In such a scenario the Kelvin-Helmholtz (KH) instability develops immediately at the edge of the beam and 
  does not require a Weibel  destabilization process to preempt it.

The KH instability is a well known instability and has been widely studied in the context  of hydrodynamic  fluid. However, 
the sheared-electron velocity flow  encountered in laboratory and astrophysical cases, mentioned above, differs from the hydrodynamic fluid flows in 
many respects.  For instance,  the sheared-flow configuration of electron fluid invariably has currents and sheared current flows associated with it.  
Consequently, the evolution of the  magnetic field  associated with it becomes an integral part of the 
 dynamics. Development of charge imbalance is another aspect in the evolution.  Though the equilibrium 
 charge balance is provided by the neutralizing static background of electrons, compressible electron flow 
 during evolution can easily lead to charge imbalance as the ions would not respond at fast electron 
 time scale phenomena. This would lead to electrostatic field generation
  which has  added influence in the dynamics. Lastly, 
  the  flow of electrons in most cases is relativistic.  Thus, to summarize 
   the KH instability in this case has additional effects due to the  presence of electromagnetic features, 
  compressibility  leading to electrostatic fields, relativistic effects etc. In the non -relativistic limit the electromagnetic effects on KH instability 
  in the context of sheared electron flows  have been investigated in detail by employing the Electron Magnetohydrodynamic (EMHD) model 
   \cite{bul,amita,amita1, Pegoraro}. This model neglects the displacement currents and space charge effects
    and assumes stationary ions which  provide the  neutralizing  background. 
   The relativistic effects on K-H instability in compressible neutral hydrodynamic fluid has been studied by Bodo $\emph{et al.}$ \cite{Bodo,Drazin}. Recently,\cite{sita} Sundar $\emph{et al.}$ have incorporated 
   relativistic effects on sheared-electron flows. This study points out crucial
   role of shear on the relativistic mass factor due to sheared velocity configuration. The 
   effect due to displacement current was retained in the relativistic regime. It was, however,  shown that for the  weakly relativistic case the effects due to 
   displacement current were negligible. However, in these studies, the space charge effects which may arise when  compressibility 
   of the electron fluid are present, have not been incorporated.
    The present article aims at 
   exploring these features using a PIC simulation. 
   
  We have carried out a 2.5D relativistic electromagnetic Particle-in-Cell simulations to study the electron shear flow instability in both cases of weak and strong  relativistic flows. By 2.5D we mean 
  that all three components  of the fields are taken into consideration, however, their spatial 
  variations are confined in 2-D plane only. 
  When the flow is weakly relativistic, we observe the development of electromagnetic 
  KH instability at the location of shear which ultimately develops into  vortices. 
  These vortices merge subsequently forming longer scales, in conformity 
  with the inverse cascade phenomena observed in typical  2-D fluid systems. The density perturbations are 
  observed to be weak in this case. The results in this case are thus very similar to the predictions 
  of the EMHD fluid behaviour. 
 When the relativistic effects are mild (and not weak), the KH instability occurs at a slower time scales. The  
 KH vortices are observed initially, which  are soon overwhelmed by compressibility effects which 
introduce magnetized non-linear  electrostatic oscillations (non-linear upper hybrid oscillations )in plasma transverse to flow. 
 In strongly relativistic regime the electrostatic oscillations dominate right from the very beginning. 
 The amplitude of the oscillations increases leading to phenomena of wave breaking. 
 In the nonlinear regime, the spectra is observed to be broad in all the three cases which implies  
 turbulence.

 The paper is organized as follows. In section II, we describe our simulation methodology.  
  The results of PIC simulations and their implications are presented in section III. It is seen that in 
  strong relativistic case compressibility effects seem to dominate resulting in electrostatic oscillations 
  transverse to the flow. These electrostatic oscillations are understood on the basis of a simplified 
  one dimensional model in section IV. Section V contains the description of the power spectrum   of the 
  fields in the nonlinear regime. 
 Section VI contains the summary and conclusions.

 \section{Description of Simulation }
\begin{figure}[1]
        \centering
                \includegraphics[width=\textwidth]{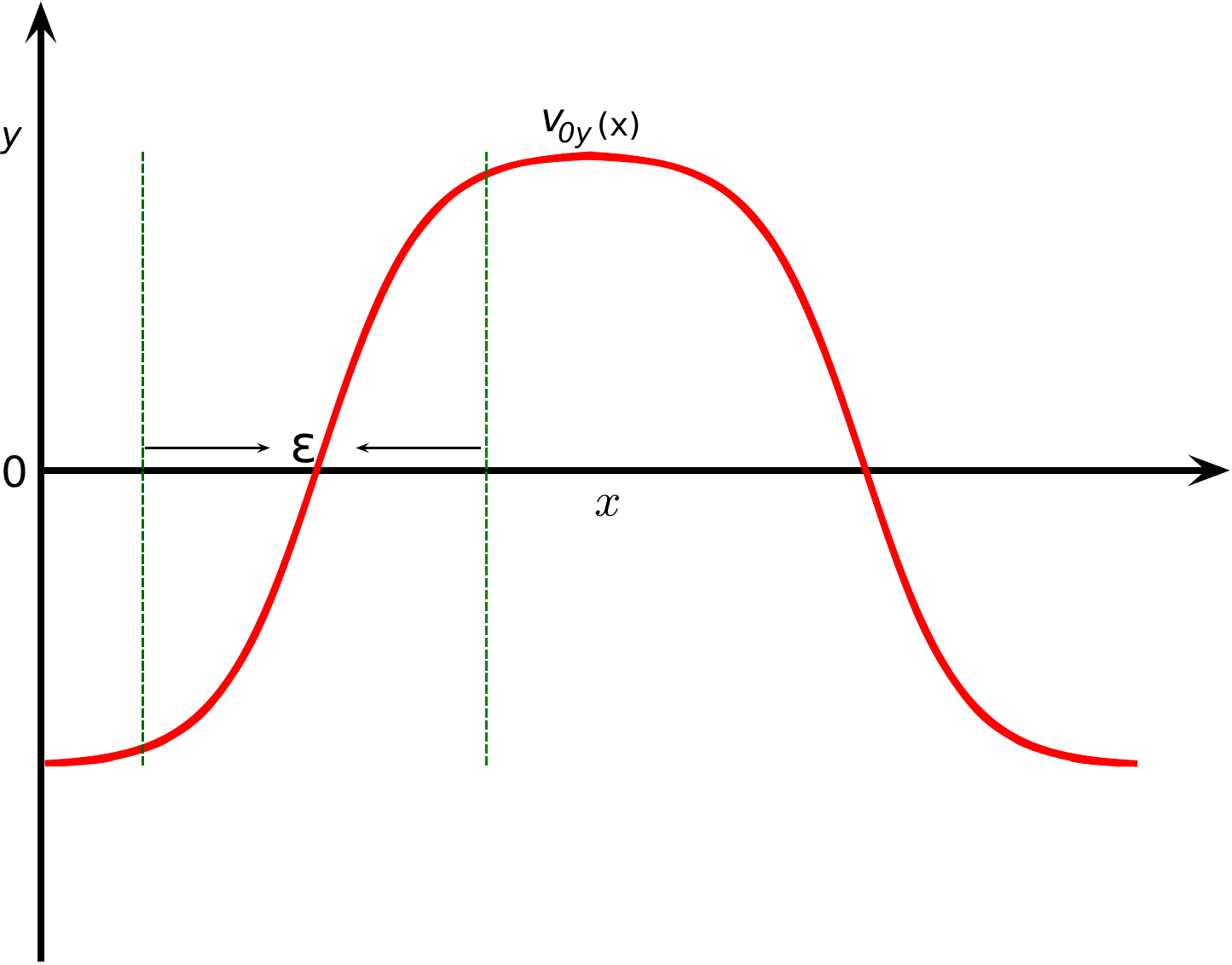} 
                 \caption {The schematic of system in present article. Initially electrons flow
                 with double tangent hyperbolic along y-axis.}
                 \label{fig:sc}
        \end{figure}

 We choose the electron to have a flow velocity along $\hat{y}$ with a  following sheared flow configuration 
 as equilibrium 
   \begin{equation}
   V_{0y}(x)=V_{0} \left[tanh((x-L_x/4)/\epsilon)+tanh((3L_x/4-x)/\epsilon)\right]-V_{0},
   \label{velocity}
  \end{equation}
  where $\epsilon$ is width of shear layer, $L_x$ is total length of simulation box in transverse direction of 
  flow and $V_{0}$ is the maximum amplitude of the flow velocity. This flow structure is shown schematically in  
  Fig.~\ref{fig:sc}. 
  The electron flow is responsible for current and produces an equilibrium magnetic field in the $B_0\hat{z}$ 
  direction. 
During the simulations, ions 
 are  kept at rest  and  merely provide for the neutralizing background. In order to satisfy the condition for 
 equilibrium force balance on electrons, there is a need to displace the electrons and ions slightly 
 in space, so that an equilibrium  electric field $\vec{E}_0$ gets created. This is chosen in such a fashion 
 so as to satisfy the condition of  
\begin{equation}
\vec{E}_0 + \frac{V_{0y}\hat{y} \times \vec{B}_0 \hat{z}}{c} = 0
\end{equation}
This ensures that the Lorentz force on electrons vanishes everywhere.  This clearly indicates the necessity 
for having an equilibrium electric field along $\hat{x}$. The electron and ion charges are thus 
displaced in an appropriate fashion so as to satisfy the 
  Gauss's law
  \begin{equation}
   \nabla.\bold E= \frac{\partial E_{x}}{\partial x}=-\frac{1}{c}\frac{\partial \left ( B_{0z}V_{0y}\right)}{\partial x}=4\pi e\left ( n_{0i}-n_{0e}\right),
  \end{equation}
here $n_{0i}$ and $n_{0e}$ are unperturbed ion and electron number densities respectively in equilibrium, e is charge of electron and c is speed of light.
To maintain equilibrium in system we have thus arranged the  electron particle number density according to following relationship \cite{coroniti},
\begin{equation}
   n_{0e}=n_{0i}+\frac{1}{4\pi ec}\frac{\partial \left ( B_{0z}V_{0y}\right)}{\partial x}.
   \label{electron_den}
  \end{equation}
  The ions are distributed uniformly with a  density $n_{0i}$ of  $3.18\times10^{18}$cm$^{-3}$ and 
 $n_{0e}$ is adjusted as per Eq.~(\ref{electron_den}).
The area of the simulation box $R$ is chosen to be $6\times5$ $( c/ \omega_{0e})^2$  corresponding to 
$600\times500$ cells; where $\omega_{0e}=\sqrt{4\pi n_{0i}e^2/m_e}$ is electron plasma frequency corresponding 
the  uniform plasma at the background  density of ions. 
 Also, $c/ \omega_{0e}= d_{e}=3.0\times 10^{-4}$cm is the skin depth.  We have used 128 particles per cell 
 for both ion and electron in our simulation. To resolve the underlying physics at the scale which is 
 smaller than the skin depth, we have chosen
a grid size  of 0.01$d_{e}$. The time step $\Delta t$, decided by the Courant condition, is 0.035 femtosecond. 

 We have considered four different set of parameters for our investigation. In all cases, velocity profile of electron is assigned by eq.~\ref{velocity}. For the first case, we choose the flow velocity of 
 electron in the weakly relativistic regime and chose the  shear width to be less than the plasma skin depth.
 We would refer this as  Case (a) which has the following parameters  $V_{0}=0.1c$, $\epsilon=0.05$ $c/\omega_{0e}$. This is the weakly relativistic case where the EMHD fluid description is supposed 
 to work pretty well.
 We consider then in case (b), the dependence of KH instability on shear width. We do this by 
 changing the value of shear width in comparison to  skin depth.  As per the EMHD description 
 the growth rate decreases when the shear width is shallow compared to the skin depth. 
 We illustrate this by specifically choosing a value of  $\epsilon=1.5$ $c/\omega_{0e}$.
 In the third and fourth cases (c) and (d) a   mild and strong relativistic limit with parameters   $V_{0}=0.5c$, 
 $\epsilon=0.05$ $c/\omega_{0e}$  and  $V_{0}=0.9c$, $\epsilon=0.05$ $c/\omega_{0e}$ are respectively 
 considered.
\section{PIC Simulation Results }
In the three subsections we discuss the results of 
(I) Weakly relativistic (II) Mild relativistic (III) Strong relativistic cases. 
 \subsection{ I. Weakly relativistic}
\begin{figure}[2]
\includegraphics[scale=0.5]{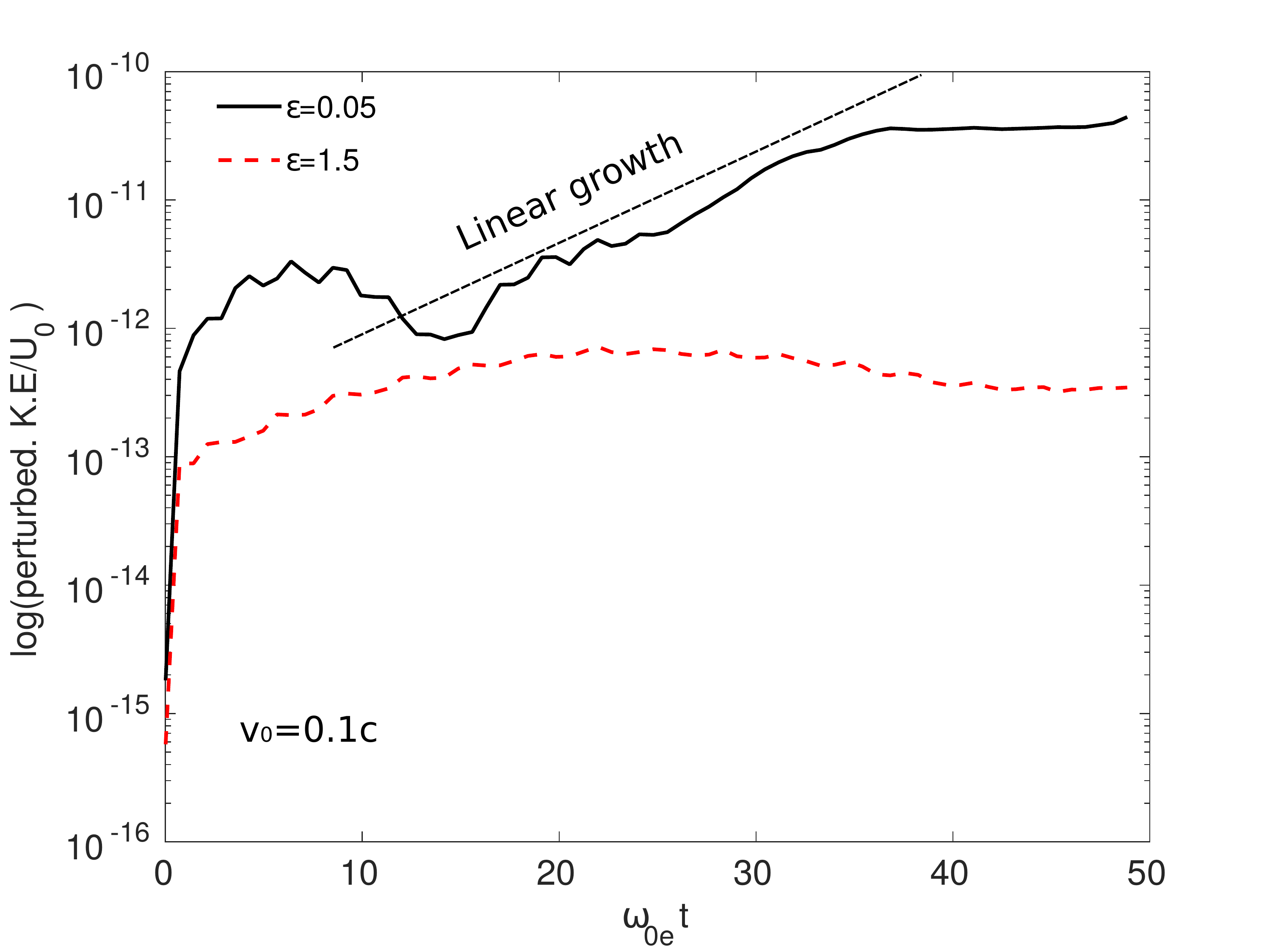} \par\bigskip\bigskip%
              \justifying   \caption{ Time evolution of perturbed kinetic energy where $u_0=(me c\omega_{0e}/e)^2$ for case (a) (black color, solid line)
              and case(b) (red color, solid line). The slope gives linear growth rate of KH instability. }  
                 \label{fig:kh1_gr}
         \end{figure}

 We choose  a the value of  $V_0 = 0.1 c$ for electron velocity to study the weakly relativistic case. 
 We observe a destabilization of the  sheared flow configuration. The  instability is tracked 
 by plotting the evolution of the 
 perturbed kinetic energy(PKE)  of the electrons in the system. This is  shown in 
Fig.~\ref{fig:kh1_gr}.  The initial steep rise is due to numerical noise. Thereafter, 
the instability grows from the  noise spectrum.  Since, the noise would lack the   
exact eigen mode structure of any particular mode, initially a combination 
of unstable modes start growing. 
 Subsequently,  as the mode with 
fastest growth dominates a linear rise in the plot of PKE can be clearly observed. 
It should be noted that evolution follows the EMHD fluid predictions of 
 the growth rate being higher for the case (a) 
when the shear width is sharper than the skin depth. In case (b) the growth is observed to be small and the 
saturation also occurs quite fast. 

\begin{figure}[!htb]
        \centering
\includegraphics[scale=0.47]{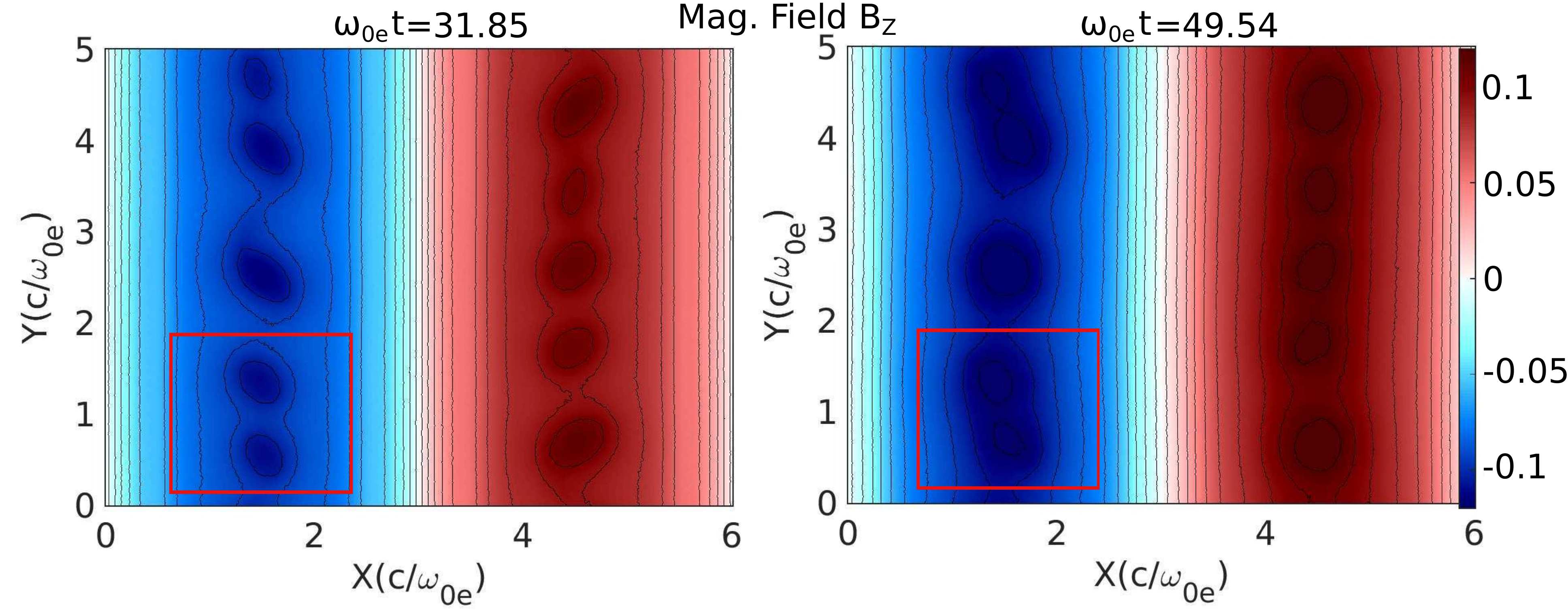} \par\bigskip\bigskip%

                    \caption { Time evolution of z-component of magnetic field $B_{Z}=B_{Z}/(me c\omega_{0e}/e)$ for case (a)  at time $ \omega_{0e}t$=31.85 and 49.54. The vortices in magnetic field are highlighted by red box (31.85$ \omega_{0e}t$)
                    which merge at later time $ \omega_{0e}t$=49.54 (highlighted by red box). }
               \label{fig:mag1}
              \end{figure}

\begin{figure}[!htb]
        \centering
\includegraphics[scale=0.465]{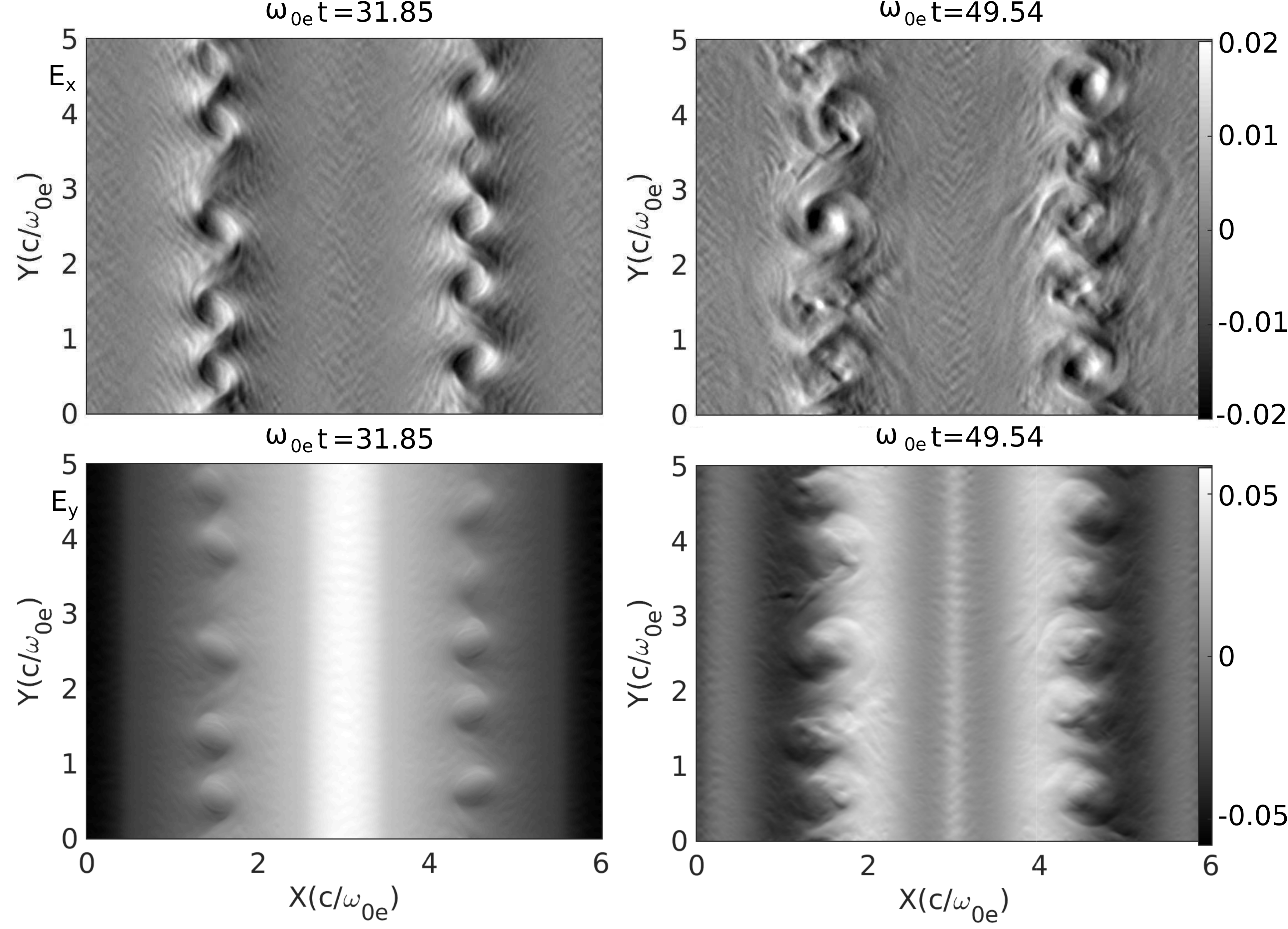} \par\bigskip\bigskip%
                    \caption { The time evolution of electric field: first row shows the x-component of electric field $E_{x}=E_{x}/(me c\omega_{0e}/e)$ and second row shows y-component of electric field $E_{y}=E_{y}/(me c\omega_{0e}/e)$ 
                     for case (a) ($V_{0}=0.1c$, $\varepsilon=0.05d_e$.)
                   }
               \label{fig:elec1}
              \end{figure}

\begin{figure}[!htb]
        \centering
\includegraphics[scale=0.465]{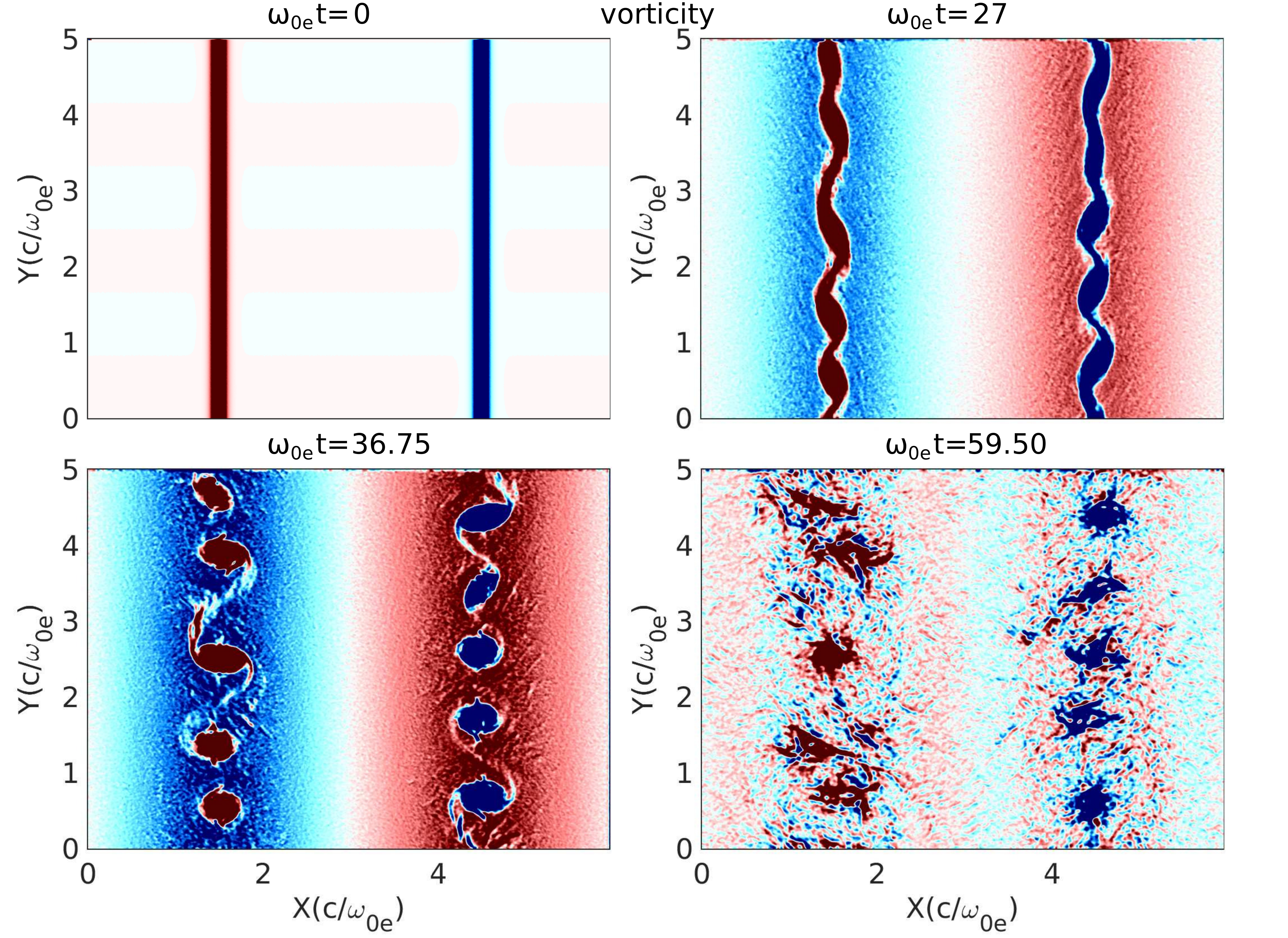} \par\bigskip\bigskip%
                 \caption { The time evolution of vorticity ($(\nabla \times V)/\omega_{0e}$) calculated from velocity field for case (a) ($V_{0}=0.1c$, $\varepsilon=0.05d_e$) which shows merging of vortices with time and turbulence stage of KH instability.}
               \label{fig:vortices}
              \end{figure}

\begin{figure}[!htb]
        \centering
        \includegraphics[scale=0.475]{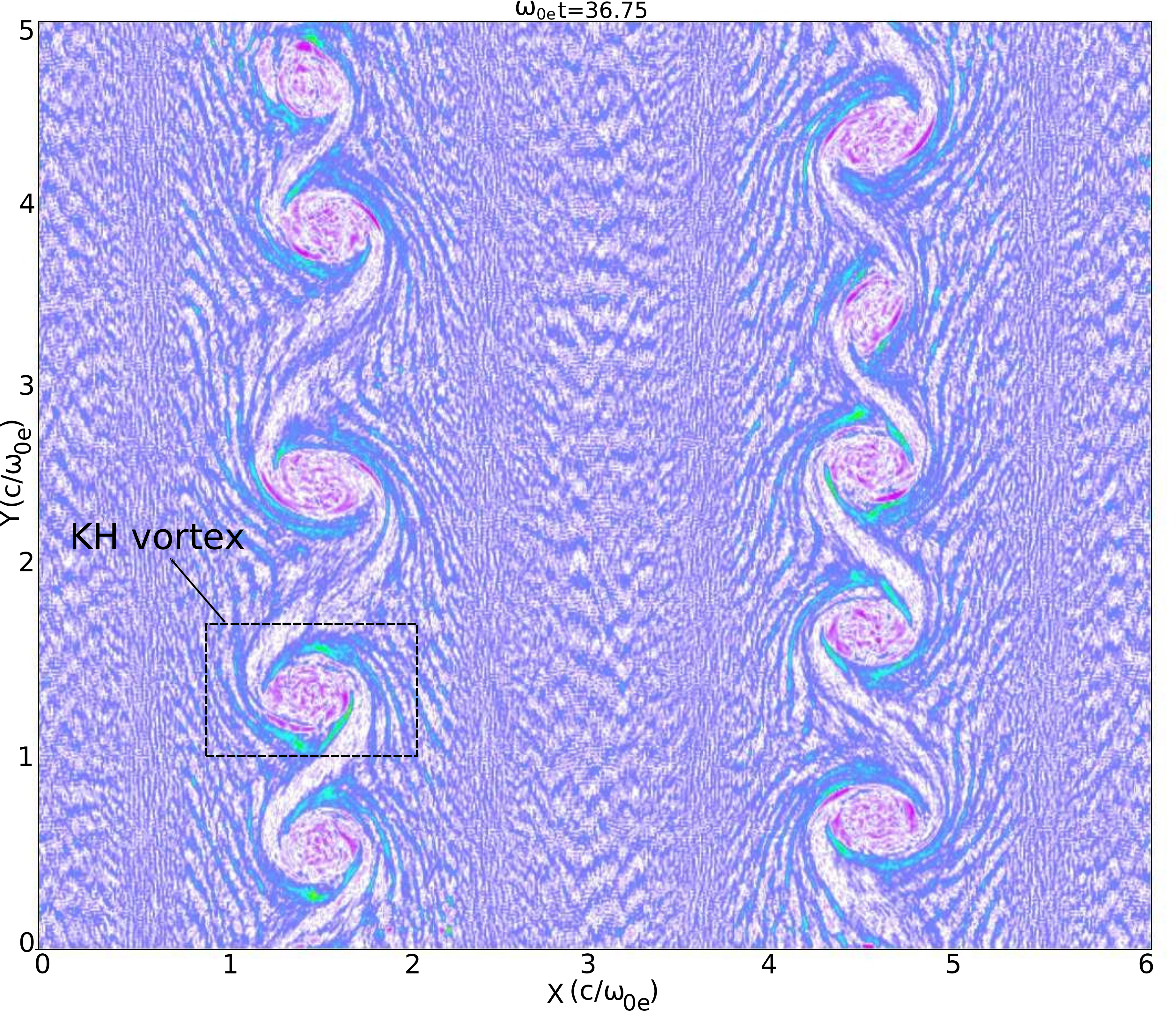} \par\bigskip\bigskip%
                 \caption { Formation of KH vortex (highlighted by black box) in electron density $n_e=n_e/n_{0i}$ at time $\omega_{0e}t=$36.75 for case (a).}
               \label{fig:kh1_density}
              \end{figure}

For a closer look at the instability development  the color contour plot of the evolution of magnetic 
field (Fig.~\ref{fig:mag1}), vorticity (Fig.~\ref{fig:vortices}) and the two components of 
Electric field (Fig.~\ref{fig:elec1}) has been shown at various times. From (Fig.~\ref{fig:mag1}) magnetic field evolution, one can 
observe that the magnetic perturbations start at the location where  velocity shear is maximum. 
These perturbations grow forming magnetic vortices which subsequently merge to form bigger structures. 
The merging process of magnetic field is along  expected lines of 2-D inverse cascade EMHD depiction of 
the problem.  The fluid vorticity also shows similar traits, however, at later times 
$t = 59.60$ (in normalized units) the long scale vorticities show signs of disintegration.  The two components of electric fields 
also show emergence of KH structures and merging. A comparison of normalized amplitudes of electric and 
magnetic field shows that the electric fields are much weaker than the magnetic fields. 
 We also show  the plot of 
electron density 
in the nonlinear regime of the KH instability at  $t = 36.75$ in Fig.~\ref{fig:kh1_density}. 
We observe that the density also acquires distinct structures of KH like vortices in the shear region. 
The density perturbations in the weakly relativistic case is observed to be weak. The maximum 
observed value of $\tilde{ne}/n_{e0} \sim 1.2$. On the other hand we would see in the strongly 
relativistic case this is as large as $8$ to $10$. 
This suggests that in the weakly relativistic regime  the instability 
essentially has an electrostatic character.


 \subsection{ II. Mild relativistic case}
\begin{figure}[!htb]
        \centering
\includegraphics[scale=0.465]{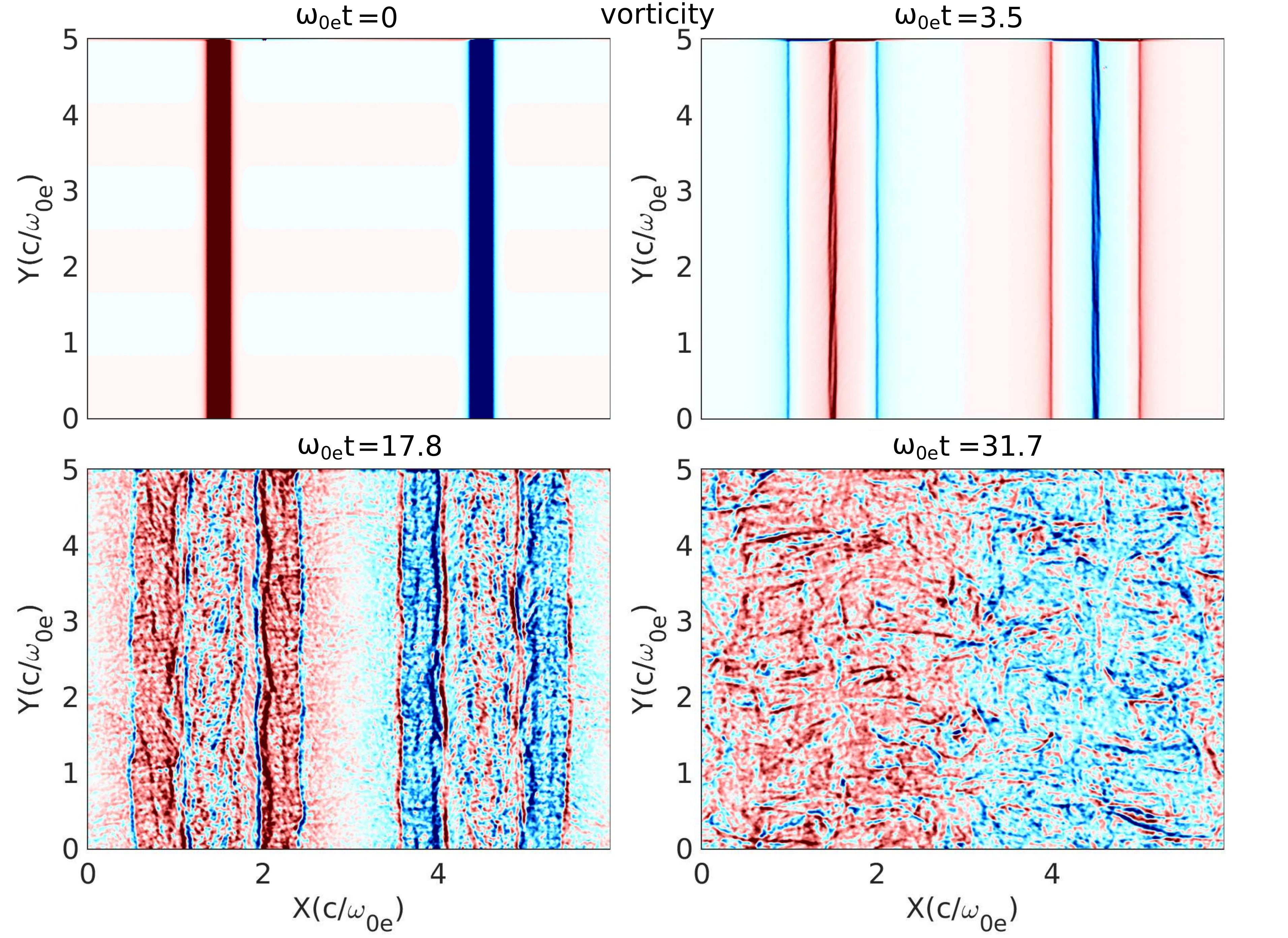} \par\bigskip\bigskip%
                 \caption { The time evolution of vorticity ($\nabla \times V$) calculated from velocity field for case (b).}
               \label{fig:vor2}
              \end{figure}

 In the mild relativistic case where $V_0 = 0.5 c$, the KH is observed to be considerably weak. 
 The vorticity plots shown in Fig.~\ref{fig:vor2} shows an initial tendency towards developing the KH rolls. 
 The KH rolls in this case are fewer in number. For case(a) they were $5$ here they are only around $3$. 
 This again suggests that the growth rate for relativistic case gets confined towards longer scale as per the predictions of EMHD model. 
 The fluid analysis carried out earlier also suggests that the cut off wavenumber of the KH moves towards 
 longer scales in mildly relativistic cases. 
 
\begin{figure}[!htb]
        \centering
\includegraphics[scale=0.465]{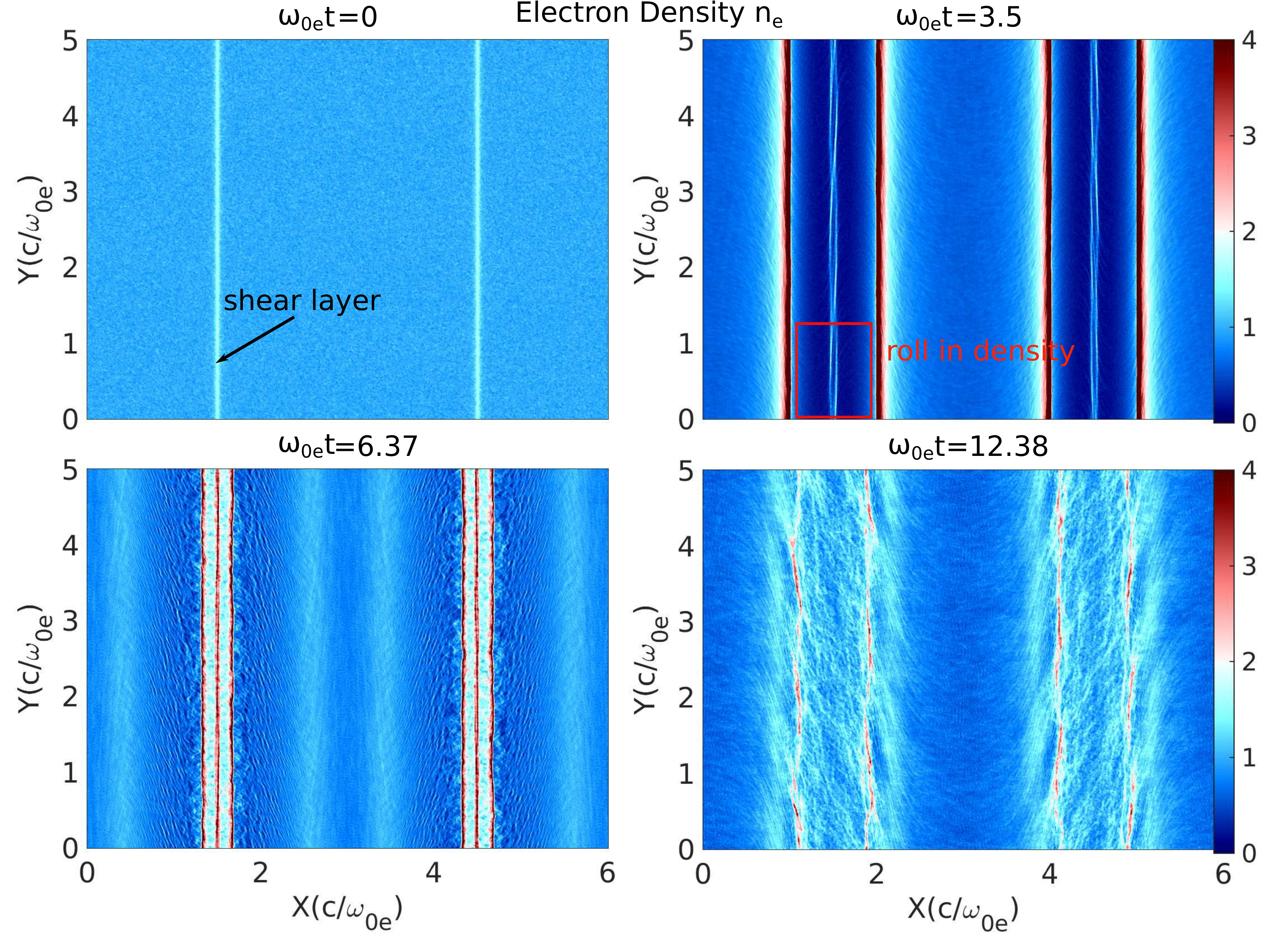} \par\bigskip\bigskip
                 \caption {The time evolution of electron density $n_e=n_e/n_{0i}$ for case (c) that shows the roll in density at time $ \omega_{0e}t$=3.5 at shear layer (highlighted by red box) which is signature of KH instability. The compression and rare faction in 
                 density can be also seen.}
               \label{fig:elec2}
              \end{figure}

 The KH rolls are observed to be  very soon overwhelmed by certain oscillations transverse to the flow. The oscillations 
 transverse to the flow are also clearly evident in the electron density plots of Fig.~\ref{fig:elec2}. 
 The density oscillations in this case are pretty strong with $\tilde{n_e}/n_{e0} \sim 4$. 
 The KH suppression and the appearance of these upper 
  electrostatic oscillations can be understood as follows. 
 As the relativistic effect increases the $\vec{V} \times \vec{B}/c$ force becomes dominating. 
 Thus a small  perturbed magnetic field $\tilde{B}$, induces  a strong $V_0\times \tilde{B}/c$ force 
  along $x$, which is responsible for the upper hybrid electrostatic oscillations.


\subsection{ III. Strong relativistic}
\begin{figure}[!htb]
        \centering
\includegraphics[scale=0.465]{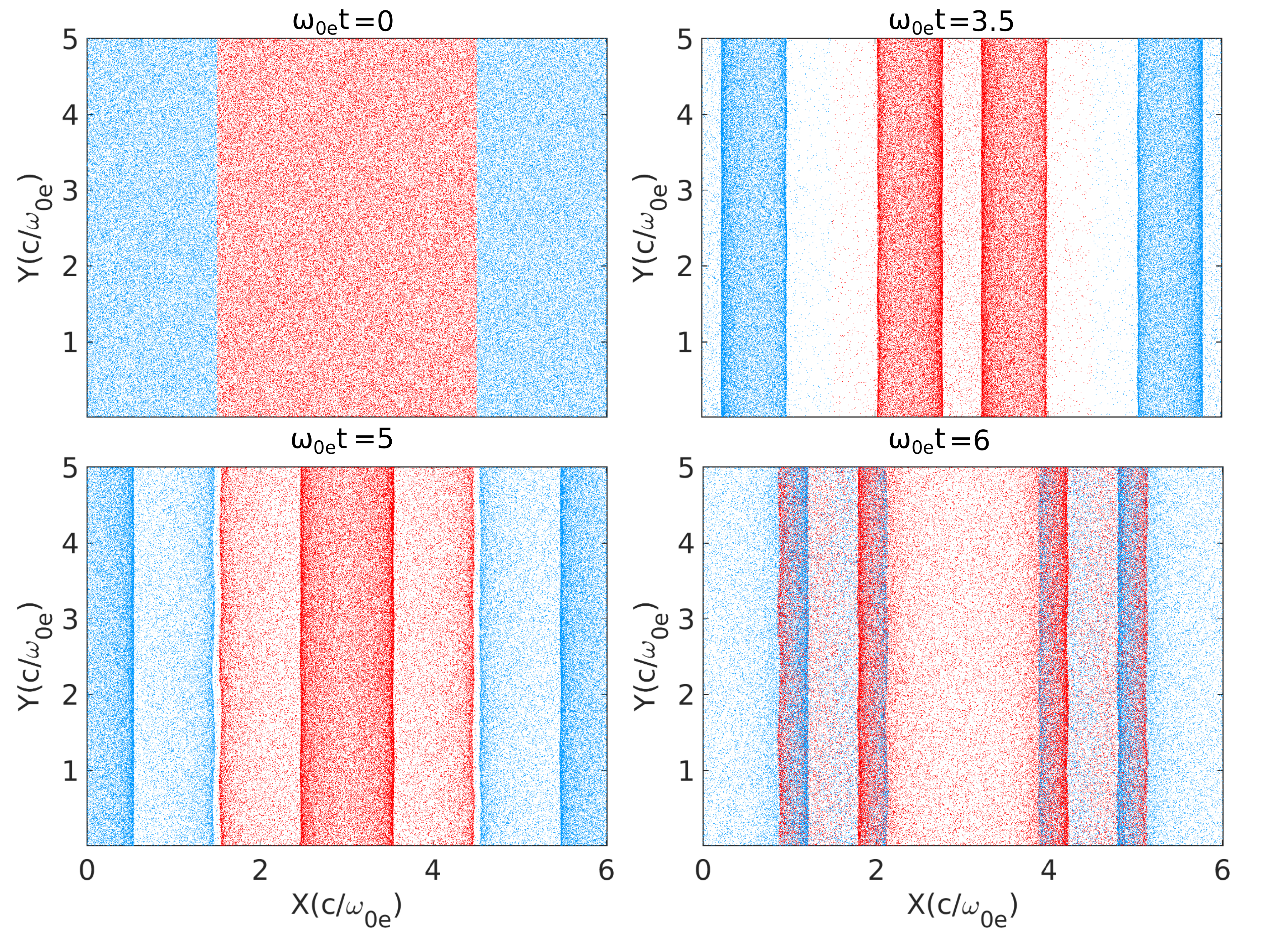} \par\bigskip\bigskip%
                 \caption { Particle picture of 2-D electron electron velocity shear configuration for case (d) which shows transverse oscillations of particles with time.}
               \label{fig:par9}
              \end{figure}

We now choose $V_0 = 0.9c$ for understanding the strongly relativistic case. 
The time evolution of PKE in this case shows the linear growth of the instability. However, the 
instability is dominated by the 
 upper hybrid electrostatic oscillations which are observed right from the very beginning. Thus the development of the 
 rolls typical of the KH instability are not very clearly evident in this case. 
Representing the initial distribution of flowing and the static electrons by different colors (red 
and blue respectively) we show the snapshots of their displacement in space in Fig.~\ref{fig:par9}. 
The electron compressibility is clearly evident, so much so and white regions 
devoid of electrons are created 
(snapshot at $\omega_{0e}t= 3.5$). The Electric fields due to background ions, however, pull 
these electrons back which results in a large amplitude  excitation of nonlinear upper hybrid electrostatic plasma 
oscillations. These oscillations are discussed in detail in the next section. 

A comparative value of the growth rate obtained from the slope of the evolution of PKE in the table below for all the cases studied by us. 
\begin{center}
{\bf{TABLE I}} \\
The maximum growth rate ($\Gamma_{gr} max.$) of K-H instability evaluated from slope of perturbed kinetic energy\\
\vspace{0.2in}
\vspace{0.2in}
\begin{tabular}{c c c c c c c c c c c c c  ll}
\hline
\hline
   &$V_{0}/c $  \hspace{0.3in}   & $ \varepsilon/(c/ \omega_{0e})$  \hspace{0.3in}        &
   $\Gamma_{max}/(V_0\omega_{0e}/c)$\\
 \hline
  &0.1    \hspace{0.3in}      & 0.05  \hspace{0.3in}          &0.7    \\
  &0.1     \hspace{0.3in}      & 1.5 \hspace{0.3in}          &0.0  \\
  &0.5    \hspace{0.3in}      & 0.05  \hspace{0.3in}       &0.34     \\
& 0.9   \hspace{0.3in}      & 0.05  \hspace{0.3in}       &0.23     \\
\hline
\end{tabular}
\end{center}
Since, classically the KH instability typically scales with the fluid flow velocity 
we have chosen to divide the growth rate with $V_0$ for a better appreciation of the comparison. 
The comparison clearly, shows that $\Gamma_{max}/V_0$ decreases due to  relativistic effects  in agreement with the earlier fluid analysis by Sundar {\it et al}. Thus the distinction between the PIC and EMHD fluid simulations finally 
boils down to the appearance and  dominance of electrostatic oscillations transverse to the flow direction. 
We study the transverse oscillations in detail in the next section. 

\section{Nonlinear upper hybrid electrostatic oscillations}
\begin{figure}[!htb]
        \centering
                 \includegraphics[trim = 0mm 0mm 0mm 0mm, clip, scale=0.35]{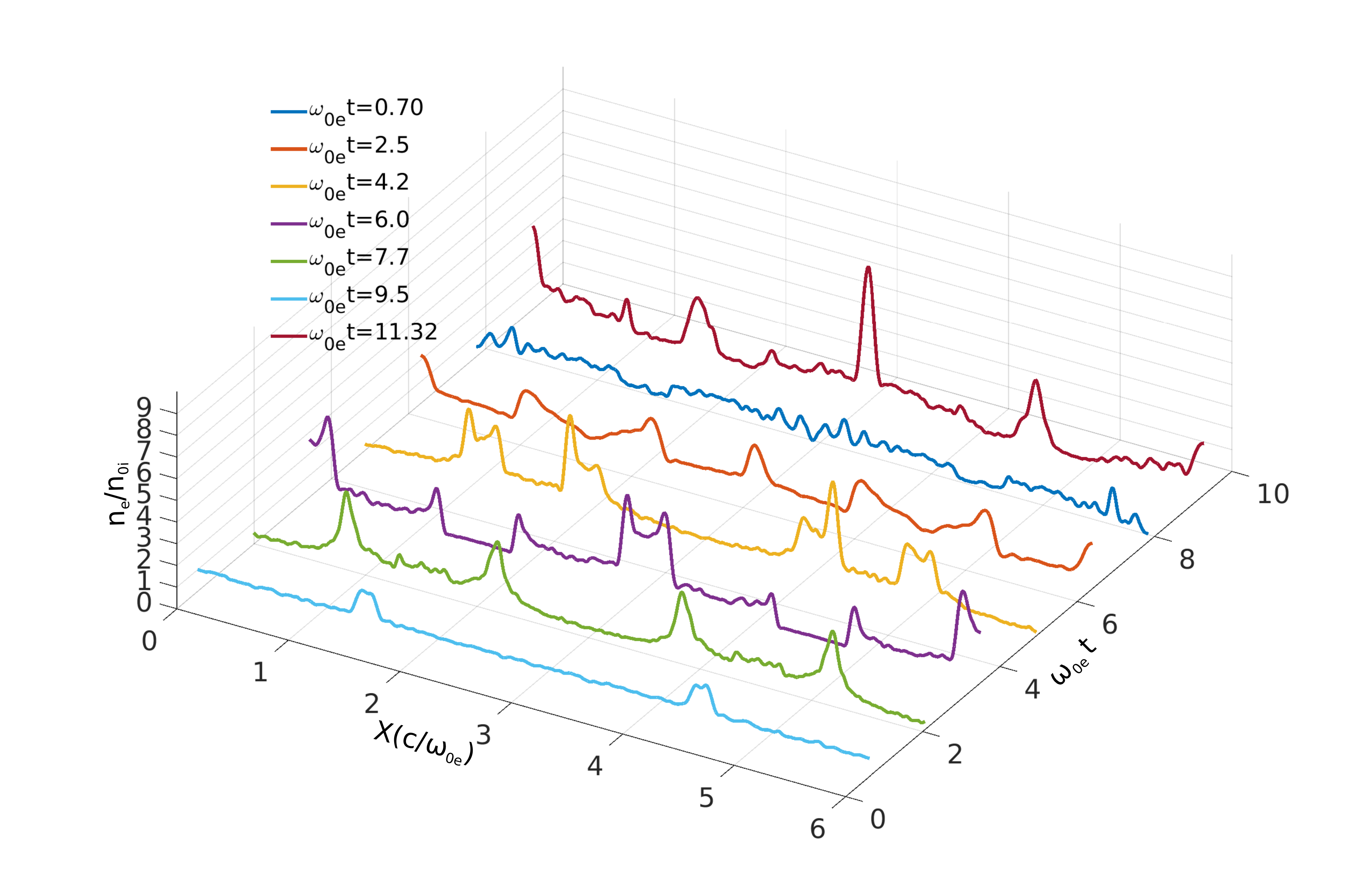}
                  \caption { Time evolution of electron density diagnosed at y $=2.5 c/\omega_{e0}$ for $V_{0}=0.9c$, $\varepsilon=0.05d_e$: This figure shows non-linear large amplitude electrostatic oscillations which break in later time.}
                   \label{fig:wavekhpic}
        \end{figure}

One of the main observations is the appearance of strong upper hybrid electrostatic oscillations 
triggered from the edge of the flow region with increasing relativistic effects. 
We show in Fig.~\ref{fig:wavekhpic} the amplitude of these oscillations as a function of time at $y = 2.5$ $c/\omega_{0e}$ for the strongly 
relativistic case of $V_0 = 0.9c$. It can be seen that the density perturbations acquire a very high 
amplitude fairly rapidly 
$\tilde{n_e}/n_{e0} \sim 8$. This is  a very nonlinear regime for the oscillations where wave breaking  
and trajectory crossing would occur. This is indeed so as the particle distribution of Fig.~\ref{fig:par9} shows clear 
crossing of blue and red electrons.

\begin{figure}[!htb]
        \centering
                 \includegraphics[trim = 0mm 0mm 0mm 0mm, clip, scale=0.65]{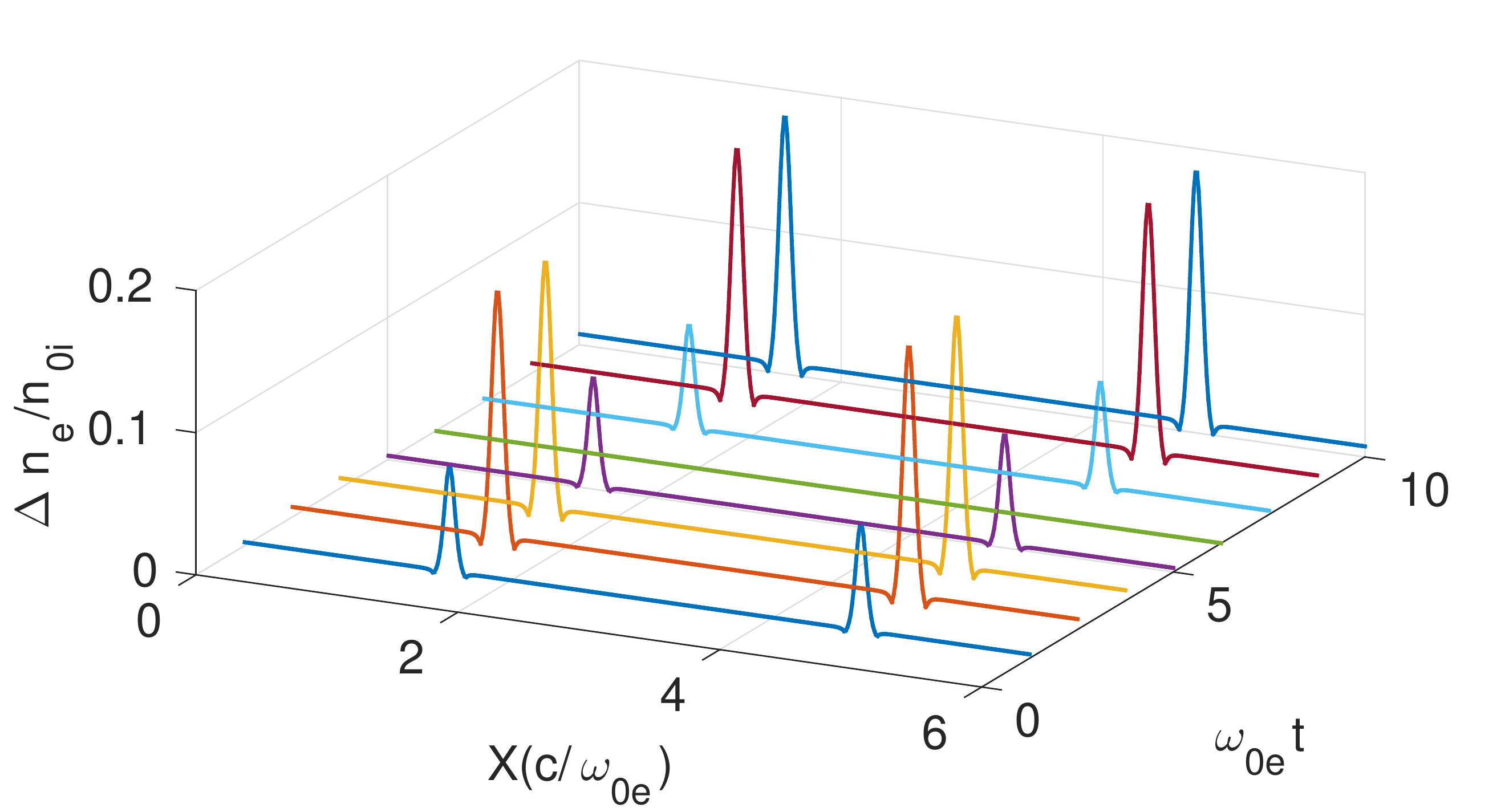}
                  \caption {Time evolution of perturbed electron density $\Delta n_e=|(n_e-n_{e0})|$ obtained from 1D model diagnosed at y $=2.5 c/\omega_{e0}$ for $V_{0}=0.1c$, $\varepsilon=0.05d_e$:This figure shows small amplitude electrostatic oscillations in presence
                  of inhomogeneous magnetic field B(x).}
                   \label{fig:wavekhnon}
        \end{figure}

\begin{figure}[!htb]
        \centering
                 \includegraphics[trim = 0mm 0mm 0mm 0mm, clip, scale=0.65]{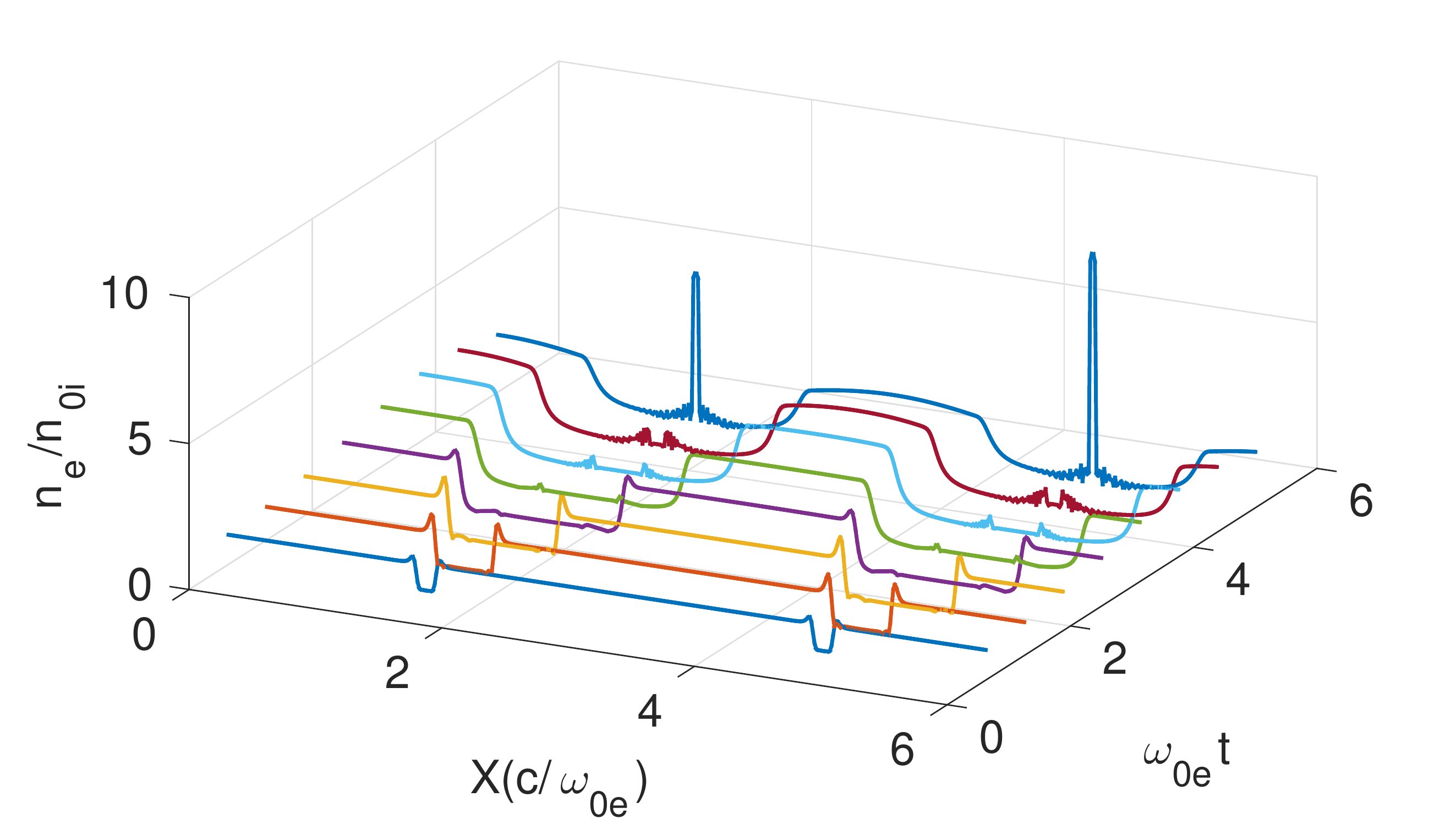}
                  \caption {Time evolution of perturbed electron density $ n_e/n_{0i}$ obtained from 1D model diagnosed at y $=2.5 c/\omega_{e0}$ for $V_{0}=0.9c$, $\varepsilon=0.05d_e$: This figure shows non-linear large amplitude electrostatic oscillations which break in later time.}
                   \label{fig:wavekhrel}
        \end{figure}

 In order to understand the dynamics behind this phenomenon, we model the phenomena by  a one-dimensional 
 magnetized  relativistic electron fluid equations for electrostatic disturbances. 
 Thus the  governing equations of the model are expressed as 
  \begin{eqnarray}
&&\left(\frac{\partial }{\partial t} +v_{ex}\frac{\partial}{\partial x} \right)n_{e}=-n_{ex}\frac{\partial v_{ex}}{\partial x}, \\
\label{con1}
&&\left(\frac{\partial }{\partial t} +v_{ex}\frac{\partial}{\partial x} \right)p_{ex}=-eE_{x}-\frac{ev_{ey}B_0(x)}{c}, \\
 \label{mom1}
&&\left(\frac{\partial }{\partial t} +v_{ex}\frac{\partial}{\partial x} \right)p_{ey}=\frac{ev_{ex}B_0(x)}{c}, \\
\label{mom2}
&&\left(\frac{\partial }{\partial t} +v_{ex}\frac{\partial}{\partial x} \right)E_x=4\pi en_{0i}v_{ex}, \\
\label{Max1}
\end{eqnarray}
 where $p_{e\alpha}=\gamma m_{e}v_{e\alpha}$ is $\alpha$-component of momentum;$\alpha$ is subscript for x and y, $\gamma=[1+p^2/m^2_ec^2]^{1/2}$ is relativistic factor and $n_{0i}$ is background ion density.
 The inhomogeneous magnetic field $B_0(x)$ is the equilibrium magnetic field generated from the 
 equilibrium electron flow considered in our PIC simulations.  For the double tangent hyperbolic profile 
 it will have the following form:  
\begin{eqnarray}
&& B_0(x)=\frac{4\pi n_{0e}e}{c}\left(V_0\epsilon\log \left(\cosh\left(0.25L_x -x\right)\right)+V_0\epsilon\log \left(\cosh\left(x-0.75L_x\right)\right)-V_0 x \right).\nonumber\\ 
 \label{B1}
\end{eqnarray}
We have solved the above equations numerically with initial profile of $v_{ey}$  using the eq.~(\ref{velocity}). For the weakly relativistic case of $V_0 = 0.1c$ the electrostatic oscillations 
that get generated are quite small and continue to remain  so indefinitely (See Fig.~\ref{fig:wavekhnon}).  
However, when the value of $V_0$ is increased to a high value of 
 $V_0=0.9c$,  large-amplitude non-linear oscillations in electron density (see Fig.~\ref{fig:wavekhrel}) 
 can be clearly seen. This is similar to the results of our PIC simulations. 
 
 The upper hybrid frequency $\omega_{UH}$ is given by
 \begin{equation}
  \omega_{UH}^2=\omega_{0e}^2+\omega_{ce}^2
  \label{uh}
 \end{equation}
 In our simulations, since the magnetic field is nonuniform, the upper hybrid oscillations occur against an inhomogeneous magnetic field background. 
 For comparing the observed oscillation frequency with that of the upper hybrid 
 oscillations we have chosen to consider an average magnetic field. Thus
 $\omega_{ce}$=$eB_{r.m.s}/m_e c$, $B_{r.m.s}$ is root mean square value of magnetic field.
 We calculate the upper hybrid frequency from dispersion relation eq.~\ref{uh}, from PIC simulation of $\omega_{UH}$ and from 1D model and have tabulated it in table II for  the two cases of mild and strong relativistic flows. 
 It can be seen that all  the two approaches (simplified dispersion equation and 1 D model) yield comparable estimates for the observed electrostatic oscillations in the PIC simulations. 
 
\begin{center}
{\bf{TABLE II}} \\
The table for upper hybrid frequency obtained from dispersion relation eq.~\ref{uh} $\omega_{UH}(anal.)$, from  PIC simulation $\omega_{UH}(PIC)$ and from 1d model $\omega_{UH}$(1d model) for various different value of $V_0$\\
\vspace{0.2in}
\vspace{0.2in}
\begin{tabular}{c c c c c c c c c c c c c  ll}
\hline
\hline
   &$V_{0}/c $  \hspace{0.3in}   & $\omega_{UH}(anal.)/\omega_{0e}$  \hspace{0.3in} & $\omega_{UH}(PIC)/\omega_{0e}$ \hspace{0.3in} & $\omega_{UH}$(1d model)$/\omega_{0e}$ 
  \\
 \hline
  &0.5    \hspace{0.3in}           &1.09    \hspace{0.3in}           &1.06  \hspace{0.3in}           &1.06    \\
  &0.9     \hspace{0.3in}          &1.27   \hspace{0.3in}          &1.26  \hspace{0.3in}          &1.25 \\
\hline
\end{tabular}
\end{center}
  The upper hybrid frequency obtained from various method are match very well and affirm the existence of upper hybrid mode in sheared electron flow. 

\section{Nonlinear regime}
\begin{figure}[!htb]
        \centering
                 \includegraphics[trim = 0mm 0mm 0mm 0mm, clip, scale=0.35]{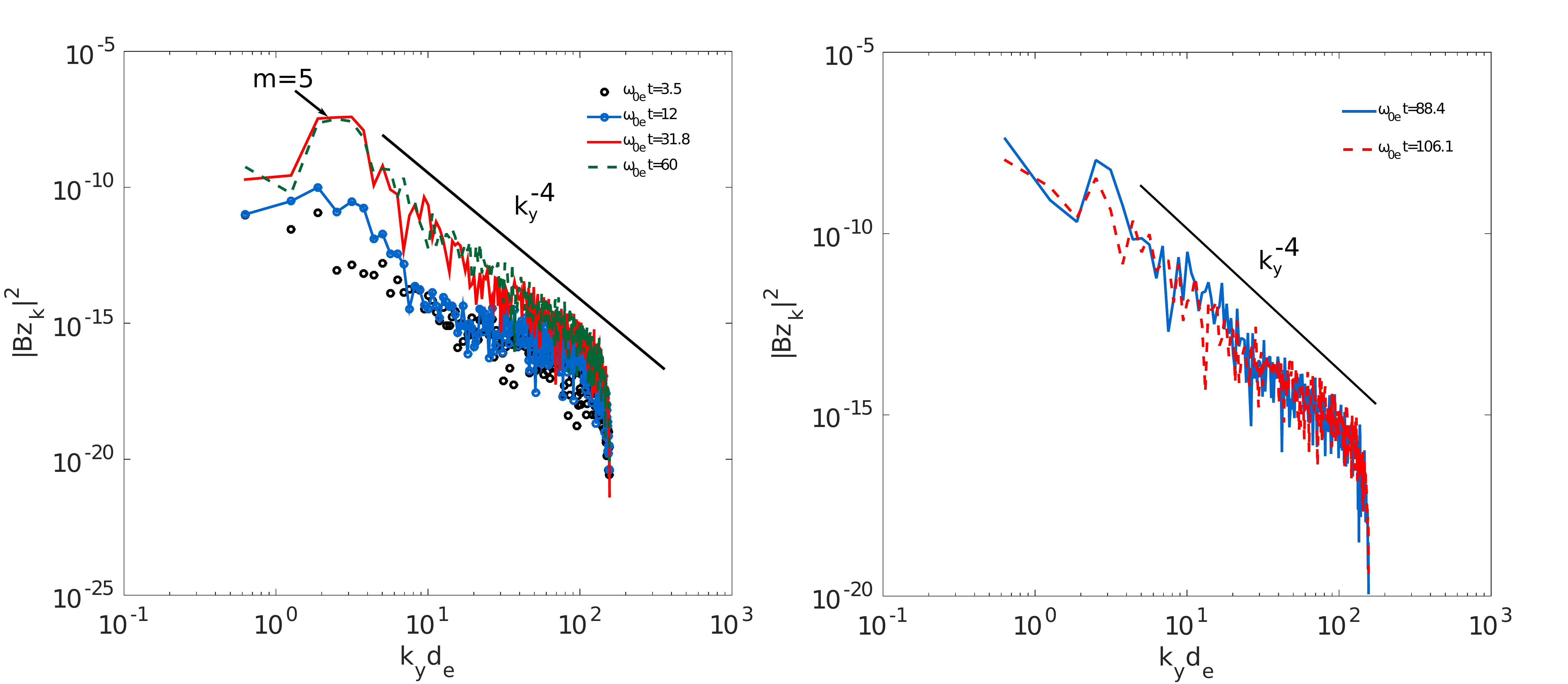}
                  \caption {Longitudinal spectrum of magnetic field energy $|Bz_{k}|^2=|Bz_{k}/(me c\omega_{0e}/e)|^2$ with time for weakly relativistic case ($V_{0}=0.1c$, $\varepsilon=0.05d_e$). 
                  (a) spectra of magnetic field up at early time where we can see the domination of power corresponding to mode m=5 (where m=$L_{y}k_y/2\pi$) (b) spectra in turbulent stage.}
                   \label{fig:spectrab1}
        \end{figure}

\begin{figure}[!htb]
        \centering
                 \includegraphics[trim = 0mm 0mm 0mm 0mm, clip, scale=0.35]{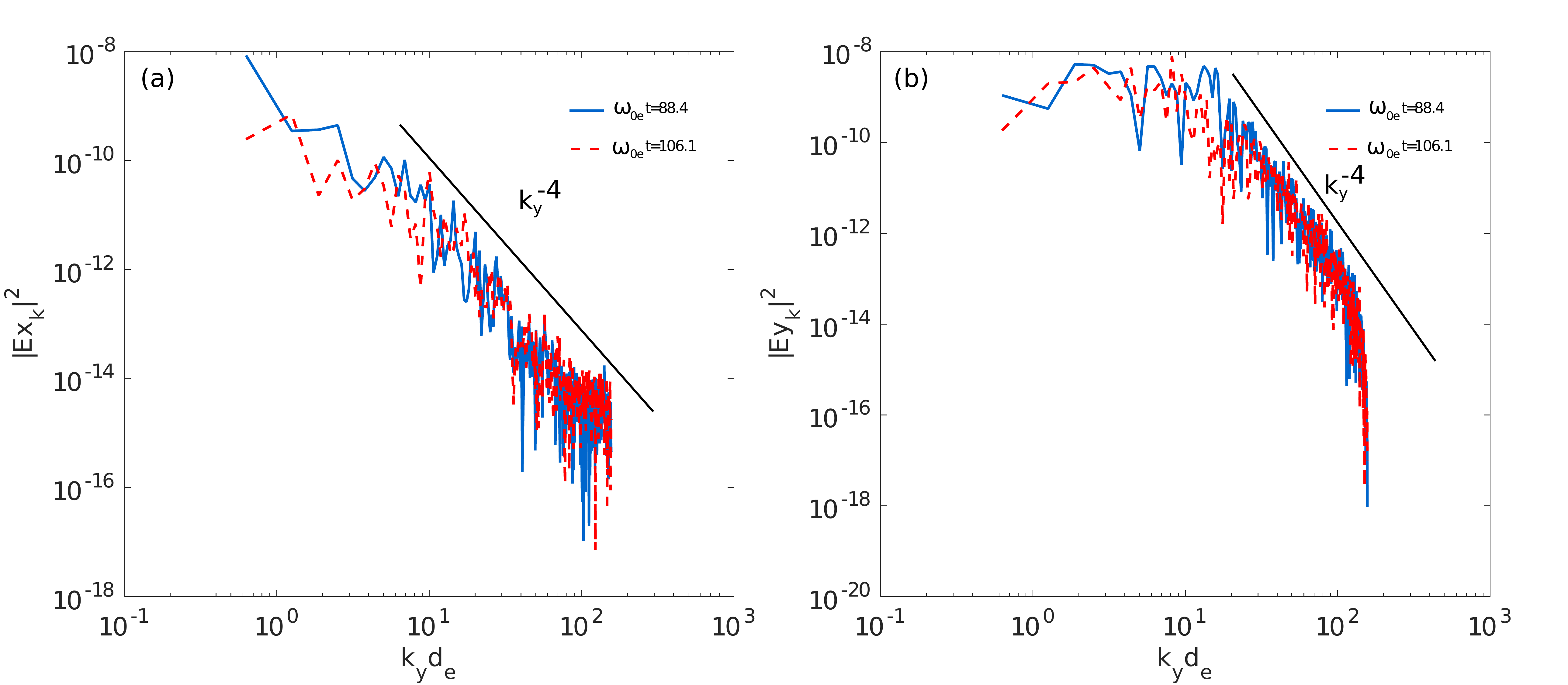}
                  \caption {Longitudinal spectrum of electric field energy with time for weakly relativistic case ($V_{0}=0.1c$, $\varepsilon=0.05d_e$). (a) spectra of x-component of electric field energy $|Ex_{k}|^2=|Ex_{k}/(me c\omega_{0e}/e)|^2$ (b) spectra of y-component of electric field energy $|Ey_{k}|^2=|Ey_{k}/(me c\omega_{0e}/e)|^2$.}
                    \label{fig:spectrae1}
        \end{figure}

\begin{figure}[!htb]
        \centering
                 \includegraphics[trim = 0mm 0mm 0mm 0mm, clip, scale=0.35]{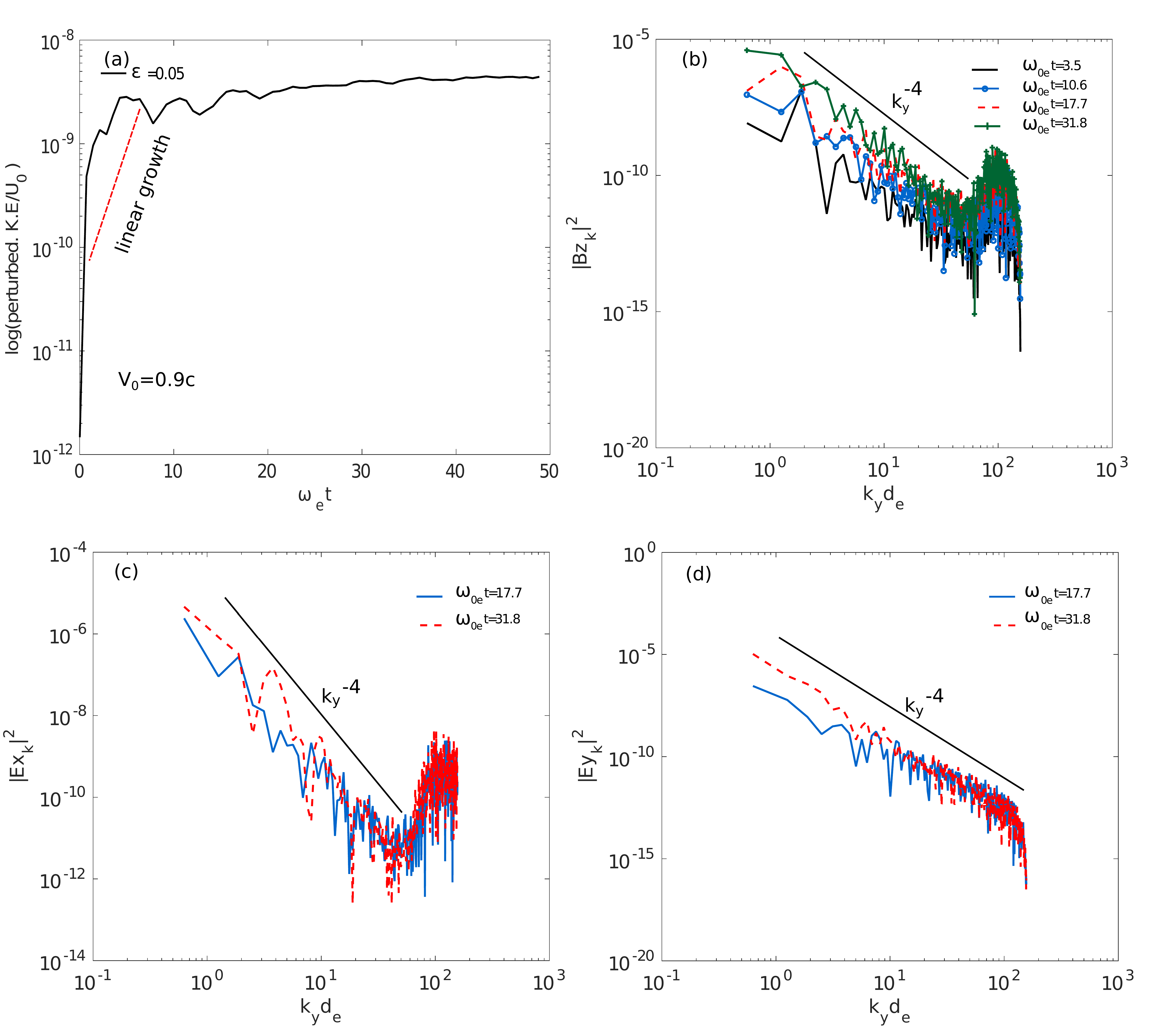}
                  \caption {perturbed kinetic energy and fields spectra for ultra-relativistic case ($V_0=0.9c$, $\varepsilon$=$0.05$ $c/\omega_{0e}$) (a) Time evolution of perturbed kinetic energy  (b) longitudinal spectra of magnetic field energy $|Bz_{k}|^2=|Bz_{k}/(me c\omega_{0e}/e)|^2$  (c) longitudinal spectra of x-component of electric field energy $|Ex_{k}|^2=|Ex_{k}/(me c\omega_{0e}/e)|^2$ 
                  (d) spectra of y-component of electric field energy $|Ey_{k}|^2=|Ey_{k}/(me c\omega_{0e}/e)|^2$.}
                   \label{fig:spectra3}
        \end{figure}

The nonlinear regime of the simulation shows evidence of turbulence generation for both 
weak and strong relativistic cases. We have plotted the spectra of magnetic and electric fields 
as a function of $k_y$ defined by the following relationship. 

\begin{equation}
   S_F(k_{y})=\frac{1}{L_{x}}\int_{0}^{L_{x}}F^2(x, k_{y}) dx,
   \label{Ey}
  \end{equation}
  where $ S_F(k_y)$ is one dimensional longitudinal energy spectra of the field F, where F is the x-
  dependent longitudinal Fourier transform of any of the electric and magnetic  fields 
  (represented by $R$ here) given by 
  \begin{equation}
  F(k_{y}, x)=\frac{1}{L_{y}}\int_{0}^{L_{y}}R(x, y)exp(-ik_y y) dy,
  \end{equation}
We observe that in both strong and weak relativistic cases the spectra is broad and has a power law 
behaviuor (see Fig.~\ref{fig:spectra3},  Fig.~\ref{fig:spectrab1} and Fig.~\ref{fig:spectrae1} ). The spectral scaling index is found to be close to $-4$. 
In the strong relativistic case, we observe that the power law extends towards the longer wavelength 
region of $kd_e \sim 1$ whereas this is not so for the weak relativistic case.    
It appears that it is easier to generate longer scales in the strongly relativistic case.

\section{Summary and Conclusion}
A detailed PIC simulation was carried out to study the instability of sheared relativistic electrons 
against a background of neutralizing ions. Our studies on weakly relativistic case show good agreement with 
the observations based on EMHD fluid approximation. For instance the observation of instability getting 
driven when the shear scale is sharper than the skin depth, the development of KH vortices in the shear 
region which ultimately merge to form longer structures etc., are all in conformity of the fluid 
EMHD theory.
In the strong relativistic case the compressibility effects are dominant and one observes a   
characteristic 
electrostatic oscillations transverse to the flow direction. This  
overwhelms the KH  in the system. The nonlinear regime in all cases 
shows a broad power spectra of magnetic field which is indicative of turbulence.



\clearpage
\newpage
\bibliographystyle{unsrt}
 
 \clearpage
\FloatBarrier
       

\end{document}